\documentclass[12pt]{spieman}  % 12pt font required by SPIE;
\usepackage{amsmath,amsfonts,amssymb}
\usepackage{graphicx}
\usepackage{setspace}

\usepackage{subfigure}
\usepackage[subfigure]{tocloft}

\title{First observation of laser-beam interaction in a dipole magnet}

\author[a,b]{Jiawei Yan}
\author[a,b]{Nanshun Huang}
\author[c,*]{Haixiao Deng}
\author[c]{Bo Liu}
\author[c]{Dong Wang}
\author[c,+]{Zhentang Zhao}
\affil[a]{Shanghai Institute of Applied Physics, Chinese Academy of Sciences, Shanghai 201800, China}
\affil[b]{University of Chinese Academy of Sciences, Beijing 100049, China}
\affil[c]{Shanghai Advanced Research Institute, Chinese Academy of Sciences, Shanghai 201210, China}

\cftpagenumbersoff{figure}
\cftpagenumbersoff{table} 
\begin{document} 
\maketitle

\begin{abstract}
As a new-generation light source, free-electron lasers (FELs) provide high-brightness x-ray pulses at the angstrom--femtosecond space and time scales. The fundamental physics behind the FEL is the interaction between an electromagnetic wave and a relativistic electron beam in an undulator, which consists of hundreds or thousands of dipole magnets with an alternating magnetic field. Here, we report the first observation of the laser-beam interaction in a pure dipole magnet, in which the electron beam energy modulation with 40-keV amplitude and 266-nm period is measured. We demonstrate that such an energy modulation can be used to launch a seeded FEL, that is, lasing at the sixth harmonic of the seed laser in a high-gain harmonic generation scheme. The results reveal the most basic process of the FEL lasing and open up a new direction for the study and exploitation of laser-beam interactions. 
\end{abstract}

% Include a list of up to six keywords after the abstract
\keywords{free-electron laser, dipole magnet, laser-beam interaction}

% Include email contact information for corresponding author
{\noindent \footnotesize\textbf{*}Haixiao Deng,  \linkable{denghaixiao@zjlab.org.cn} }

{\noindent \footnotesize\textbf{+}Zhentang Zhao,  \linkable{zhentangzhao@zjlab.org.cn} }
\begin{spacing}{2}   % use double spacing for rest of manuscript

\section{Introduction}
\label{sect:intro}  % \label{} allows reference to this section
A charged particle radiating energy in the form of an electromagnetic wave when it is accelerated is the basic principle behind modern accelerator-based light sources. Among such sources, synchrotron radiation and free-electron lasers (FELs) have played key roles in numerous scientific fields by providing high-brightness electromagnetic waves over a wide spectral range. X-ray FELs \cite{Pellegrinireview,HUANG2021100097}, which are considered to be the next generation of light sources, are capable of providing femtosecond X-ray pulses with a peak brightness ten orders of magnitude higher than the third-generation synchrotron light sources. Compared to synchrotron radiation, the amplification of the FEL pulse comes from the strong and continuous interaction between an electromagnetic wave and a relativistic electron beam in a periodic lattice of alternating dipole magnetic fields, known as an undulator.

In an FEL process, the interaction between the electromagnetic wave and electrons causes energy modulation of the electron beam. The energy modulation evolves into longitudinal density modulation on the scale of the FEL wavelength, referred to as bunching. The bunching contributes to the FEL power growth, and the amplified FEL power further enhances and speeds up the bunching. This positive feedback loop leads to an exponential growth of the FEL power in a high-gain FEL. The amplification of the FEL eventually saturates when the bunching reaches a maximum. In the over-saturated regime, the energy exchange will also work in an undesired way, i.e., energy is extracted from the photon field to the electron beam. Owing to the existence of such a feedback mechanism, the undulator spontaneous emission can be used as an initial seed to drive the FEL amplification at short wavelengths, a mechanism known as self-amplified spontaneous emission, which is the baseline principle of most X-ray FELs worldwide \cite{lcls,sacla,decking2020mhz,pal,swissfel}.

In addition to the FEL power amplification, the presence of energy modulation caused by laser-beam interaction also enables modern X-ray FEL facilities to achieve better performance and various unique characteristics \cite{hemsing2014beam}. Seeded FELs employ an external laser to trigger the FEL frequency up-conversion, making them ideal for providing stable, fully coherent X-rays \cite{HUANG2021100097}. Typically, in a seeded FEL scheme, such as high-gain harmonic generation (HGHG) \cite{yu1991generation}, the energy of the electron beam is periodically modulated by a seed laser. A dispersive magnetic chicane is used to transform the purely sinusoidal energy modulation into density bunching that contains high harmonic components. Then, the bunched electron beams are used to produce highly coherent FEL pulses at the preferred harmonics. More advanced seeded FEL schemes are being developed to improve the frequency multiplication efficiency \cite{PhysRevLett.102.074801,PhysRevLett.111.084801,PhysRevSTAB.16.010702}. In addition, a laser heater \cite{PhysRevSTAB.7.074401,PhysRevLett.124.134801} is normally placed before the bunch compressor in the linac section of an X-ray FEL facility, where an external laser is employed to interact with the electron beam, thereby increasing the uncorrelated beam energy spread and thus suppressing the microbunching instability induced by the effects of the longitudinal space charge and coherent synchrotron radiation. Furthermore, the laser-beam interaction is proposed to widely manipulate and rearrange the electron distributions and then generate attosecond coherent pulses \cite{PhysRevLett.92.224801,PhysRevSTAB.8.040701,PhysRevSTAB.9.050702,PhysRevLett.110.084801, PhysRevAccelBeams.19.080702, PhysRevLett.126.104802}, mode-locked X-ray pulse sequences \cite{PhysRevLett.100.203901,Kur_2011}, and X-rays with orbital angular momentum \cite{PhysRevLett.102.174801,PhysRevLett.106.164803} to satisfy different scientific requirements.

\begin{figure*}[htp] 
	\centering 
	\includegraphics[width=0.85\linewidth]{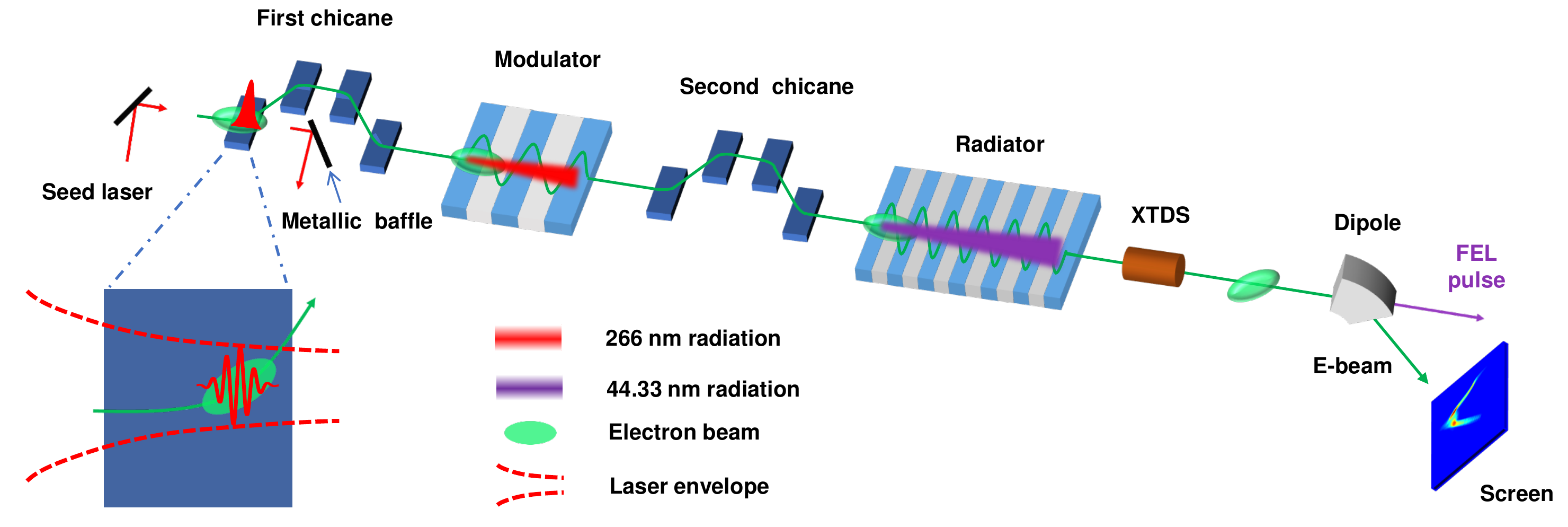}
	\caption{Schematic layout of the experiment. An 800-MeV electron beam is sent to the first chicane and interacts with a 266-nm seed laser in the first dipole magnet. Energy modulation and density modulation are performed simultaneously in the first chicane. In the modulator, an electron beam is used to generate coherent radiation at the fundamental wavelength, which also enhances the energy modulation. The radiator is used to produce FEL pulses at the sixth harmonic of the seed laser. }
	\label{scheme01} 
\end{figure*}

An electron beam moving along a straight line cannot efficiently interact with an electromagnetic wave in free space. The beam needs a curved trajectory to achieve efficient interaction with the electromagnetic field. The energy change of the electron with a horizontal motion in an electromagnetic field is given by $mc^{2}\frac{\mathrm{d} \gamma}{\mathrm{d} t}=\mathrm{e}v_{x}E_{x},$ where $\gamma mc^{2}$ is the electron energy, $E_{x}$ is the horizontal electric field of the electromagnetic wave, and $v_{x}$ is the horizontal velocity of the electron. Conventionally, one undulator period is treated as a standard unit in FEL physics; for example, the well-known FEL equations \cite{BONIFACIO1984373} and well-benchmarked codes GENESIS \cite{reiche1999genesis} and GINGER \cite{fawley2002user} are based on the undulator-period-averaged mode. In fact, the electrons experience a magnetic field with sinusoidal variations in intensity during an undulator period, which can be considered as a series of dipole magnetic fields of different intensities. In principle, the interaction between the electromagnetic wave and the electron beam has already taken place when the electron beam passes through each of the basic components of the undulator, namely a dipole magnetic field, which is the fundamental process of the FEL interaction \cite{DENG2010508}. However, to the best of our knowledge, all the FEL interactions between the electromagnetic wave and the electron beam are accomplished and experimentally observed in a full undulator, i.e., tens to hundreds or even thousands of dipole magnetic fields. In this work, we designed and conducted an experiment at the Shanghai soft X-ray FEL (SXFEL) test facility \cite{zhao2017status} to study the interaction between an electromagnetic wave and an electron beam in a pure dipole magnetic field. In the experiment, the energy modulation of the electron beam induced by an ultraviolet laser in a dipole magnet was observed and measured. Furthermore, the feasibility of seeded FEL lasing using such energy modulation was also demonstrated. 

\section{Experimental characterization of laser-beam interaction}

The schematic layout of the experiment is shown in Fig.\ \ref{scheme01}. An 800-MeV electron beam with 1 ps (FWHM) bunch length, 600 pC bunch charge, and 1.5 mm-mrad normalized transverse emittance was sent into the first chicane, which consisted of four rectangular dipole magnets with an effective length of 0.3 m. The width of these dipole magnets is 0.25 m. A 266-nm external laser with a pulse length of 160 fs (FWHM) was expected to interact with the electron beam at the first dipole magnet, and a metallic baffle was placed in the middle of the chicane to draw the laser out so that the electron beam did not interact with the laser at the fourth dipole magnet. Furthermore, it should be noted that there is an injection chicane located 7.7 m before the first chicane for laser injection. In the injection chicane, the laser and the electron beam could interact in the fourth dipole magnet with a length of 0.2 m. In the experiment, the laser waist was tuned to near the first dipole magnet of the first chicane, and thus the interaction in the injection chicane could be ignored. The rms radius of the laser at the first dipole was around 200 $\rm \mu m$. The transverse envelope size of the electron beam was  also focused at around 200 $\rm \mu m$. Once the beam energy modulation is formed in the first dipole magnet, it is converted to density bunching when the electron beam passes through the remaining part of the entire chicane. This means that, in this experiment, both energy modulation and density bunching were accomplished in the first chicane. In addition to the intensity and profile of the laser, the energy exchange obtained from the laser-beam interaction also depends on the electron beam trajectory in the laser field. Therefore, the interaction in the dipole magnet was correlated with the strength of the magnetic field. Because the magnetic field of the dipole magnet also determines the dispersion strength of the chicane, the energy modulation and density bunching were mutually coupled in the chicane.

\begin{figure}[htp] 
	\centering 
	\includegraphics[width=0.6\linewidth]{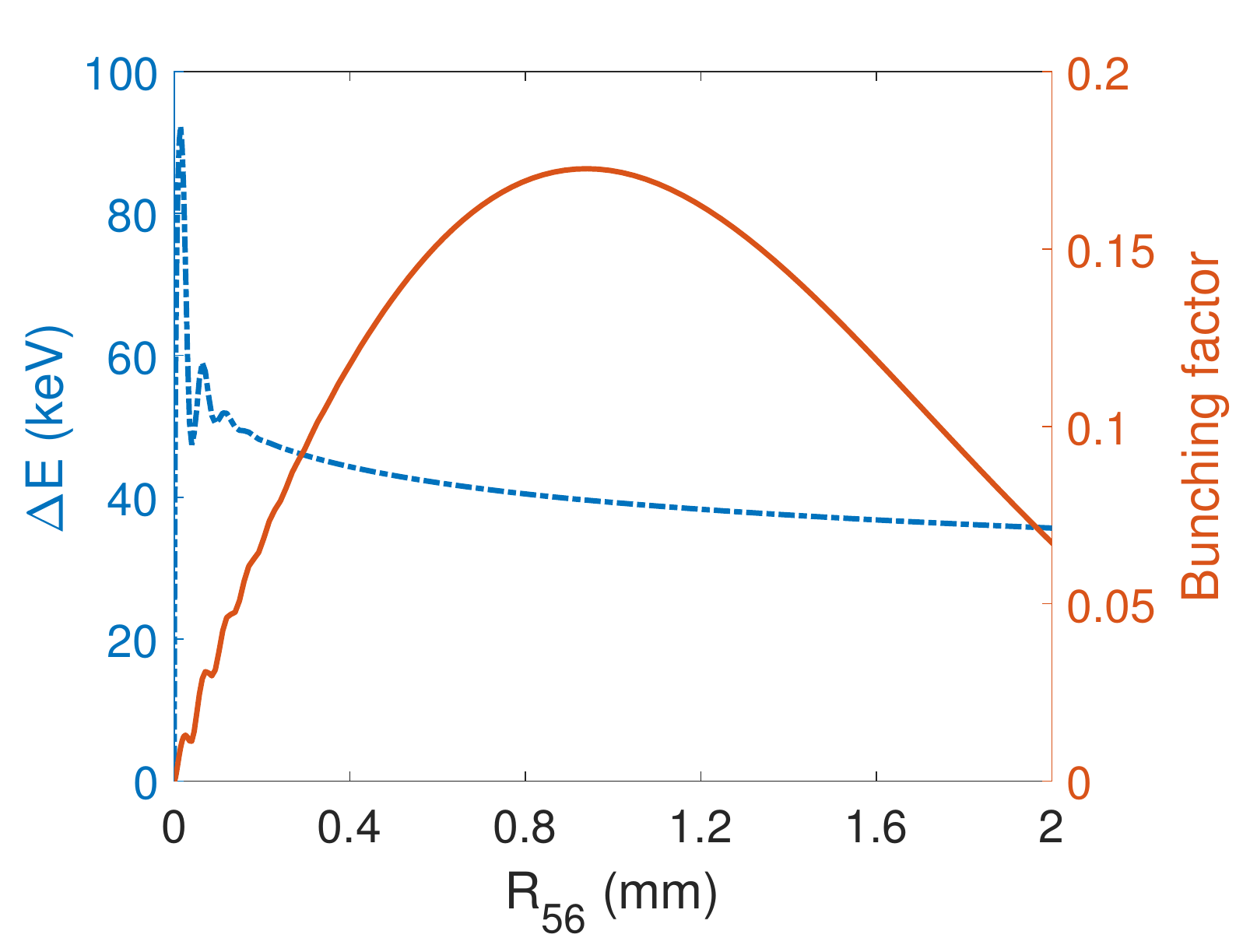}
	\caption{Three-dimensional tracking of the laser-beam interaction. Plotted are the energy modulation amplitude of the electron beam (dashed line) and bunching factor (solid line) as a function of the dispersion strength of the first magnetic chicane.}
	\label{theory} 
\end{figure}

\begin{figure*}[!htb]
	\centering
	\includegraphics[width=0.85\linewidth]{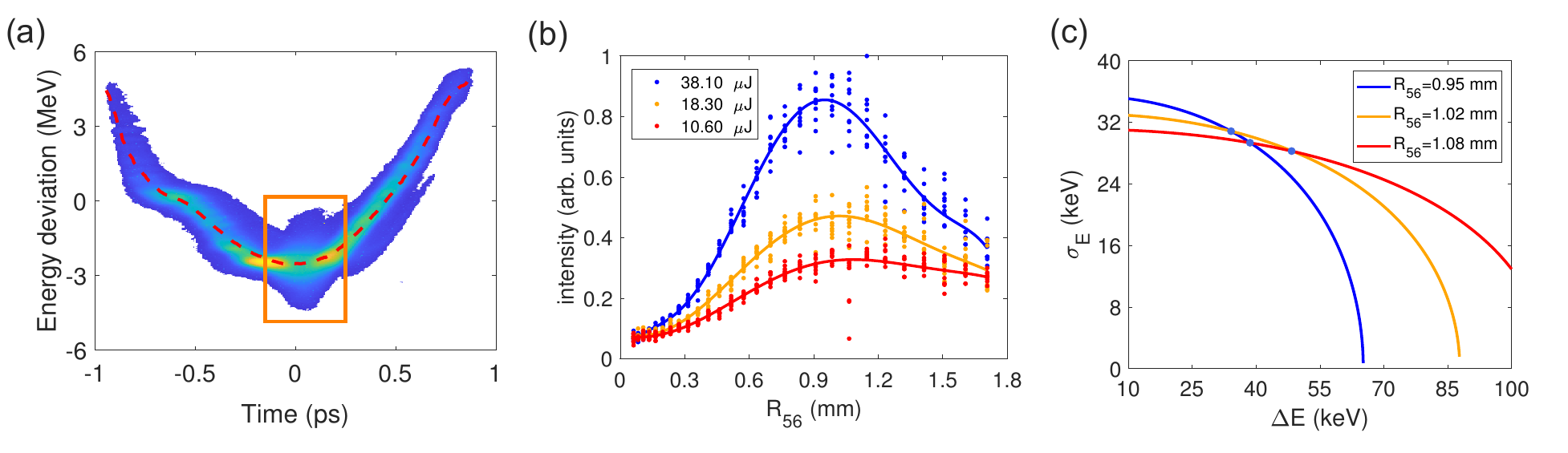}

	\caption{Experimental characterization of the laser-beam interaction in the dipole magnet. (a) Longitudinal phase space of the electron beam after the interaction. The red dashed line represents the central energy of the electron beam. The orange box contains areas that are altered due to the laser-beam interaction. Beam head is on the left. (b) Measured coherent radiation intensity and fitted curves after the electron beam passes through the first chicane under different laser pulse energies. The coherent radiation intensity is recorded by a photodiode when the dispersion strength of the first chicane is scanned. The three optimal $R_{56}$ obtained  were 0.95, 1.02, and 1.08 mm when the seed laser pulse energies were 38.10, 18.30, and 10.60 $\rm \mu J$, respectively. (c) Calculation results of the energy modulation amplitude and the initial slice energy spread using the coherent radiation generation method \cite{PhysRevSTAB.14.090701}. }
	\label{r561scan}
\end{figure*}

We first used a three-dimensional tracking algorithm based on electrodynamics \cite{DENG2010508} to analyze the laser-beam interaction in the dipole magnet. In the simulation, all parameters were determined according to the experimental settings. The initial slice energy spread of the electron beam was set to 30 keV according to the measurements in the normal operation of the SXFEL. The peak power of the laser was set to 220 MW. The density modulation of the electron beam was quantified by the bunching factor \cite{yu1991generation}. As presented in Fig.\ \ref{theory}, the simulation reveals the relationship between the value of $R_{56}$ of the chicane and the obtained energy modulation amplitude as well as the bunching factor at the fundamental wavelength after passing through the chicane. When $R_{56}$ is set to 0.01 mm, an energy modulation amplitude of 92 keV can be obtained. However, owing to constraints from the actual chicane configuration, the dipole magnetic field of the first chicane is adjustable from 0.03 to 0.12 T, corresponding to $\rm R_{56}$ values of 0.11 to 1.7 mm. In this range, the energy modulation changes slowly from 52 to 36 keV, and a bunching factor greater than 0.1 can be obtained at the fundamental wavelength. In theory, an electron beam with such a strong bunching factor can be used to produce significant coherent radiation.

In the experiment, the electron beam was sent to interact with the external laser in the first dipole magnet of the first chicane. The laser pulse energy was set to be its maximum, i.e., 38.10 $\rm \mu J$. According to the three-dimensional tracking result, the $R_{56}$ value of the first chicane was set at around 0.9 mm, corresponding to a dipole magnetic field strength of 0.086 T. To observe the changes in the electron beam after the interaction, the gaps of the modulator and the radiator were fully opened. Here, we used an X-band transverse deflecting structure (XTDS) and a dipole magnet at the end of the undulator section to measure the longitudinal phase space of the electron beam. When the laser pulse was successfully synchronized with the electron beam, significant changes in the phase space were observed in the electron beam, and the location of these changes shifted with the adjustment of the laser timing synchronization. Fig.\ \ref{r561scan}(a) shows a typical longitudinal phase space of the electron beam with changes in the central part.

We further measured the energy modulation amplitude induced in the dipole magnetic field through the generation of coherent radiation \cite{PhysRevSTAB.14.090701}. In the measurement, the following modulator with a length of 1.5 m and a period of 80 mm was set to be resonant at 266 nm and thus to generate coherent radiation at the fundamental wavelength of the external laser. Next, we scanned the dispersion strength of the first chicane to determine the optimal dispersion strength that maximizes the intensity of the coherent radiation. As presented in the three-dimensional simulation results, the energy modulation amplitude varies very slowly in the scanning range of $R_{56}$. Therefore, we considered that the energy modulation amplitude was fixed under a specific external laser. According to the measurement method, a numerical relationship between the energy modulation amplitude and the average slice energy spread can be obtained through the optimal dispersion strength of the chicane. Consequently, multiple numerical relationships need to be obtained by varying the laser pulse energy, and their intersections were treated as the measurement result. In addition to 38.10 $\rm \mu J$, two other laser pulse energies of 18.30 and 10.60 $\rm \mu J$ were also used to interact with the electron beam. The measured intensities of the coherent radiation and the fitted curves based on the measurement under the three laser pulse energies are shown in Fig.\ \ref{r561scan}(b). Three optimal $R_{56}$ values of 0.95, 1.02, and 1.08 mm were obtained under the three different laser pulse energies. As displayed in Fig.\ \ref{r561scan}(c), according to the three optimal $R_{56}$ values, three numerical relationships are modeled, where all the energy modulation amplitudes are scaled to that obtained by the laser pulse energy of 38.10 $\rm \mu J$. The average of the three intersections was regarded as the measurement result. The measured slice energy spread was 29 keV, which is consistent with the measurement results under normal FEL operation. The measured energy modulation amplitude induced by the external laser with 266-nm wavelength and 38.10-$\rm \mu J$ pulse energy in a dipole magnet of 0.3 m was 40 keV, which reasonably agrees with the simulation result. 

\section{Lasing feasibility of a seeded FEL}

We have demonstrated and measured the laser–beam interaction in a dipole magnetic field. The obtained energy modulation is a factor of 1.4 greater than the slice energy spread in our case. Since the dipole magnet and laser pulse coexist in many parts of an FEL facility and the energy modulation introduced can be considerable, laser-beam interactions in the dipole magnetic field cannot be ignored and need to be carefully considered in the design and operation of the FEL, especially in cases where the ultimate output performances are pursued. In addition, the energy modulation induced in a dipole magnet has many potential applications. Because a laser-induced energy spread of 20 keV is sufficient to suppress microbunching instabilities, a chicane without an undulator could be a future option for the laser heater. Moreover, even larger energy modulation would be obtained by using a laser with a higher peak power or an optimized dipole magnet, thus generating a fully coherent FEL by exploiting energy modulation is a more straightforward application. 

\begin{figure}[htp] 
	
	\centering 	
	\includegraphics[width=0.85\linewidth]{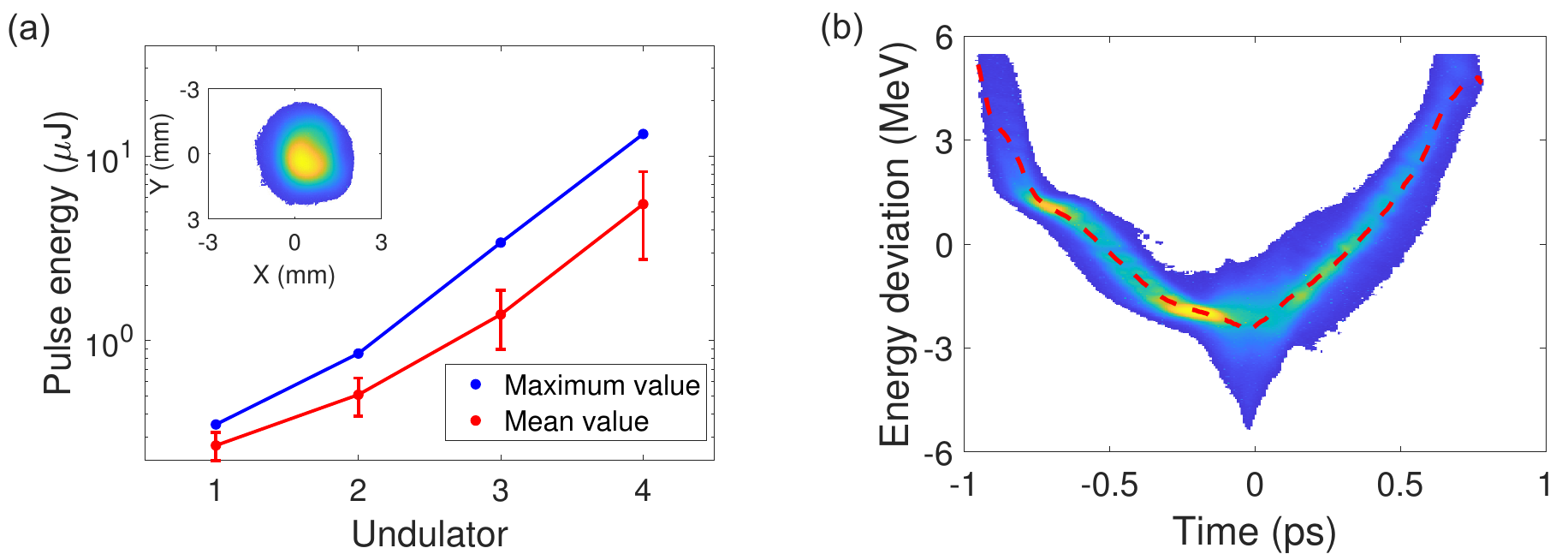}
	\caption{Performance of the FEL lasing at the sixth harmonic of the seed laser. (a) Gain curve and typical transverse profile of the FEL pulse at 44.33 nm. The red and blue points represent the average pulse energy and maximum pulse energy at the end of various undulator segments, respectively. The FEL pulse energy was measured by a calibrated photodiode at the end of the undulator section. The error bars represent the root-mean-square intensity fluctuations. The inset displays one typical transverse profile of the FEL pulse. (b) Typical longitudinal phase space of the electron beam at the exit of the radiator.}
	\label{Gain} 
\end{figure}

In this work, we explored the feasibility of using the energy modulation obtained in the dipole magnet for FEL lasing at the sixth harmonic of the seed laser. Limited by existing hardware, it is difficult to increase the peak power of the seed laser or change the dipole magnet. The energy modulation amplitude of a factor of 1.4 greater than the slice energy spread  is too weak to lase at the sixth harmonic directly; consequently, the self-modulation method \cite{yan2020selfamplification} was employed to further enhance the energy modulation. In the experiment, the pulse energy of the seed laser was kept at 38.10 $\rm \mu J$. The dispersion strength of the first chicane was set at the optimal value to maximize the intensity of the coherent radiation in the modulator. As a result, the electron beam was modulated by the coherent radiation generated by itself, and the energy modulation was enhanced. Thereafter, we sent the electron beam to the radiator, which was composed of four undulator segments with a length of 3 m and a period of 40 mm. Initially, only the first undulator segment was used for coherent radiation generation at the sixth harmonic of the seed laser. Then, the dispersion strength of the second chicane was scanned to maximize the coherent radiation intensity. According to the optimal dispersion strength of 0.16 mm obtained from the scanning and the previously obtained slice energy spread of 29 keV, we estimated an energy modulation of 247 keV after self-modulation. The results indicated that the energy modulation amplitude was increased sixfold through self-modulation.

\begin{figure}[htb]
	\centering
	\includegraphics[width=0.85\linewidth]{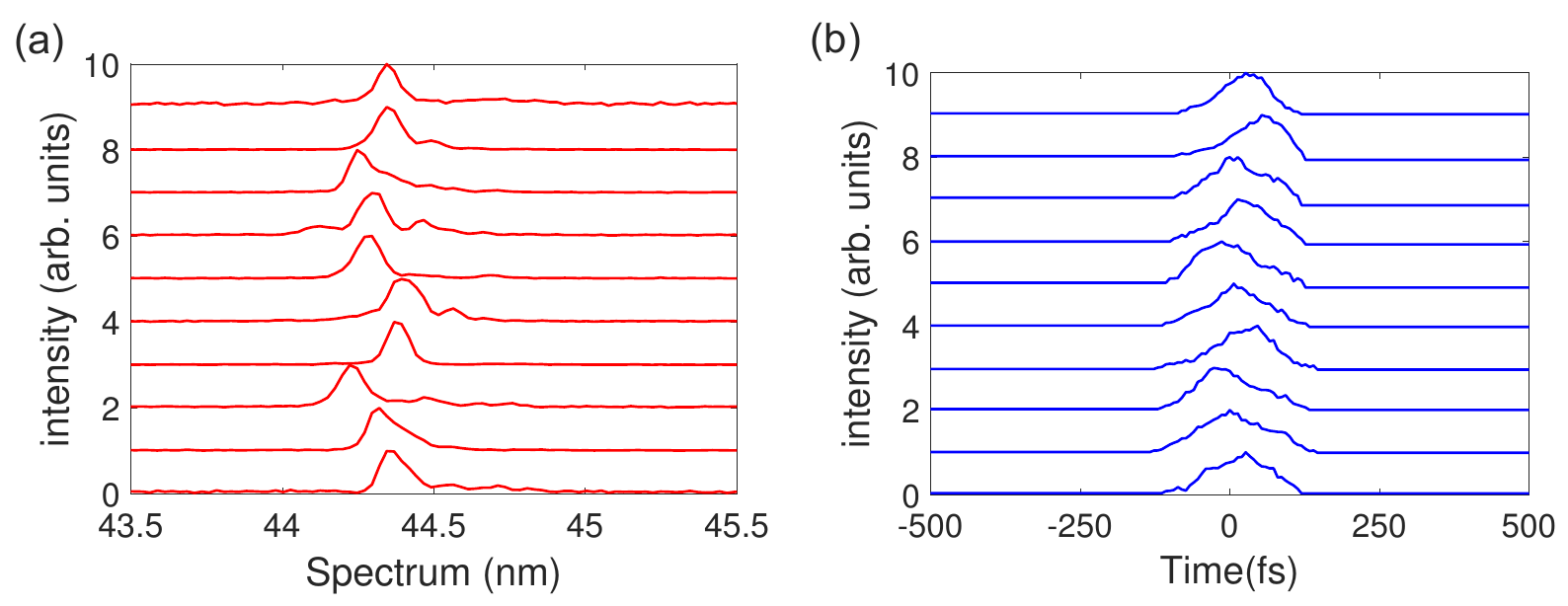}
	\caption{Spectra and reconstructed temporal profiles of 44.33-nm FEL pulses. (a) Ten typical spectra measured by the spectrometer after the radiator. (b) Power profiles of ten typical FEL pulses reconstructed by using the XTDS system.}
	\label{tp}
\end{figure}

Combined with the enhanced energy modulation after the modulator, the second chicane and the radiator were used for HGHG lasing at 44.33 nm. The dispersion strength of the second chicane was maintained at the optimal value. Other undulator segments of the radiator were used to further amplify the 44.33-nm radiation. Fig.\ \ref{Gain}(a) presents the maximum and average pulse energies of the FEL pulse after each undulator segment and one typical transverse profile of the FEL pulse. At the end of the radiator, FEL pulses with an average pulse energy of 5.5 $\rm \mu J$ and a rms energy jitter of 2.7 $\rm \mu J$ were obtained. One typical longitudinal phase space of the electron beam after the radiator is shown in Fig.\ \ref{Gain}(b). A drop in the centroid kinetic energy can be observed at the central part of the electron beam. The spectra of the FEL pulses were measured using a spectrometer with a resolution of 0.05 nm. The average FWHM bandwidth over 60 consecutive shots was $\rm 1.7 \times10^{-3}$. Fig.\ \ref{tp}(a) shows ten typical measured spectra. In addition, we reconstructed the temporal profile of the FEL pulse according to the change in the centroid energy of the electron beam \cite{ISI:000335224400003}. Ten typical temporal profiles of FEL pulses are presented in Fig.\ \ref{tp}(b). The average pulse length (FWHM) of the ten pulses is 113 fs. The obtained temporal profiles and spectra are similar to those obtained during normal operation using undulators. Moreover, larger pulse energies can be expected through careful optimization of the electron beam orbit and focusing. Shorter wavelengths can be obtained by using cascaded HGHG operation.

\section{Discussion}

In summary, for the first time, we have demonstrated and measured the interaction between a laser and relativistic electrons in a pure dipole magnetic field. A 266-nm laser was used to interact with an 800-MeV electron beam in a dipole magnet. The measured energy modulation amplitude was 40 keV, which is consistent with the theoretical tracking results. This experiment intuitively reveals a fundamental step in the FEL process and thus contributes to a deeper understanding of FEL physics. Moreover, as an example, we have shown that the energy modulation obtained in a dipole magnet can be used for lasing at the sixth harmonic of the seed laser in an HGHG setup.

The results presented here open a new direction for the study and exploitation of the interaction between lasers and relativistic electrons. A simple dipole magnet can be used to introduce energy modulation of relativistic electrons, thus effectively suppressing microbunching instabilities or tailoring the FEL pulse properties. Using a seed laser with a peak power of hundreds of gigawatts, it is possible to directly obtain an energy modulation amplitude on the order of MeV for seeded FELs without additional self-modulation. Furthermore, based on this scheme, the complex manipulation of the electron beam in echo-enabled harmonic generation \cite{PhysRevLett.102.074801} can be accomplished in two magnetic chicanes, which greatly simplifies the configuration of the facility. Because the laser-beam interaction in a dipole magnet does not have a specific resonance condition, it is well tolerant of energy jitter and chirp and is promising for electron beams from plasma wakefield based accelerators. In addition, the results also enlighten the design of novel radiators with a non-sinusoidal magnetic field.

From another perspective, our results highlight that the existence of laser-beam interaction in a dipole magnetic field may lead to considerable energy modulation. As more and more laser systems are used in FEL facilities, this effect should be taken into account in many FEL operations. In particular, using an external laser to modulate an electron beam normally requires the use of a chicane for laser injection, which means that the laser inevitably interacts with the fourth dipole magnet of the chicane. For FEL schemes that utilize laser modulation for attosecond pulse generation, an external laser with high intensity is often employed. In such cases, the laser-beam interaction in the laser injection chicane, or even in the correctors, would significantly perturb the initial beam quality, which may degrade the final achieved FEL performance. Therefore, the energy modulation occurring in the dipole magnets should be reasonably matched to the modulation in the undulator.

\section*{Authors’ contributions}
J.W.Y. and H.X.D. proposed, designed, and realized the experiment at the SXFEL test facility. J.W.Y., H.X.D., and Z.T.Z co-wrote the manuscript. All authors contributed to the construction and commissioning of the SXFEL.

\section*{Acknowledgments}
	This work was partially supported by the National Key Research and Development Program of China (Grant Numbers 2018YFE0103100), and the National Natural Science Foundation of China (Grant Numbers 11935020). The authors thank Zihan Zhu, Hanxiang Yang, and Weilun Qin for fruitful discussions.

\section*{Data availability}

The data that support the findings of this study are available from the corresponding author upon reasonable request.

%%%%% References %%%%%

\bibliography{report}   % bibliography data in report.bib
\bibliographystyle{spiejour}   % makes bibtex use spiejour.bst

%%%%% Biographies of authors %%%%%

\vspace{2ex}\noindent\textbf{Jiawei Yan} is currently a physicist at the European XFEL. He received his PhD in 2021 from the Shanghai Institute of Applied Physics, Chinese Academy of Sciences, under the supervision of Professor Haixiao Deng. His research interests include free-electron laser physics, electron beam dynamics, and machine learning.

\vspace{2ex}\noindent\textbf{Nanshun Huang} is currently a PhD student with Professor Haixiao Deng at the Shanghai Institute of Applied Physics, Chinese Academy of Sciences. He received his B.S. degree from the University of South China in 2017. Now his research interests are focused on X-ray free-electron laser oscillators.

\vspace{2ex}\noindent\textbf{Haixiao Deng} is currently a professor at the Shanghai Advanced Research Institute, Chinese Academy of Sciences. He received his PhD from the Shanghai Institute of Applied Physics, Chinese Academy of Sciences, in 2009. His research interest focuses on particle accelerator based light sources.

\vspace{2ex}\noindent\textbf{Bo Liu} is currently a professor at the Shanghai Advanced Research Institute, Chinese Academy of Sciences. He received his PhD from the Institute of High Energy Physics, Chinese Academy of Sciences, in 2005. His research focuses on free-electron laser, laser technology, and synchronization.

\vspace{2ex}\noindent\textbf{Dong Wang} is currently a professor at the Shanghai Advanced Research Institute, Chinese Academy of Sciences. He received his PhD from the Institute of High Energy Physics, Chinese Academy of Sciences, in 1998. His research interests include accelerator physics and free-electron laser physics.

\vspace{2ex}\noindent\textbf{Zhentang Zhao}  is currently a professor at the Shanghai Advanced Research Institute, Chinese Academy of Sciences. He received his PhD from the Tsinghua University in 1990. He is an academician of the Chinese Academy of Engineering. His research mainly focuses on synchrotron light source, free-electron laser, and proton therapy facility.

%\vspace{1ex}
%\noindent Biographies and photographs of the other authors are not available.
%
%\listoffigures
%\listoftables

\end{spacing}
\end{document}